\documentclass[prb,preprint]{revtex4}

\pdfoutput=1

\usepackage{graphicx}

\begin{document}


\title{Stripes in heterovalent-metal doped cuprates}

\medskip 

\date{July, 4 2021} \bigskip

\author{Manfred Bucher \\}
\affiliation{\text{\textnormal{Physics Department, California State University,}} \textnormal{Fresno,}
\textnormal{Fresno, California 93740-8031} \\}

\begin{abstract}
Doping $La_2CuO_{4}$ with alkaline-earth, $Ae= Sr, Ba$, (possibly co-doped with lanthanide $Ln = Nd, Eu$) generates holes in the $CuO_2$ planes of $La_{2-z-x}Ln_zAe_xCuO_{4}$ crystals. Pairs of the holes turn $O^{2-}$ ions to neutral oxygen atoms. A fraction of the atoms, $\tilde{O}$, of hole density $\check{p} \le  0.02$, is itinerant, skirmishing 3D-AFM to suppression at $\check{p}$.
The remaining oxygen atoms, $O$, are stationary at anion lattice sites and form a superlattice that gives rise to both charge-order stripes and magnetization stripes with incommensurability $q_{c,m}(x) \propto \sqrt{x-\check{p}}$ for $Ae$-doping up to a watershed value, $x \le \hat{x}$. More doping causes overflow of new holes to the $LaO$ layers, leaving stripes of constant $q_c$ in the $CuO_2$ planes. Antiparallel orientation of magnetic moments $\mathbf{m}(O)$ yields a natural explanation for the coupling of $q_m(x) = \frac{1}{2} q_c(x)$. Hole population in the $LaO$ layers may also be responsible for the watershed in the doping dependence of X-ray intensity, diffracted by stripes, upon cooling through the transition to superconductivity. 
Doping $Ln_2CuO_{4}$ ($Ln$ = $Nd, Pr, La$) with tetravalent $Ce$ generates electrons that reside pairwise in copper atoms. The large $Cu$ atom size may be the reason for the structural T $\rightarrow$ T' transition. The magnetic moments $\mathbf{m}(Cu) \;[ \simeq \mathbf{m}(Cu^{2+})]$ align with the AFM of the host. This causes the high 3D-AFM stability in the $n$-doped compounds as well as the lack of magnetization stripes. 
When $n$-doped, $q_c(x) \propto \sqrt{x}$.
Above a threshold temperature $T'$, electron-hole pairs are thermally generated, but then separate to reside pairwise at $Cu$ and $O$ atoms. The latter, adding to the $Ae$-generated holes, account for the increase of $q_c(x,T)$ with temperature. By aligning with the AFM of the host, the magnetic moments of thermally generated $Cu$ atoms counteract the magnetic moments from the $O$ atoms.  This breaks the locking of the incommensurability of charge-order and magnetization stripes, $q_m(x) \ne \frac{1}{2} q_c(x)$.

\end{abstract}

\maketitle

\pagebreak
\includegraphics[width=5.51in]{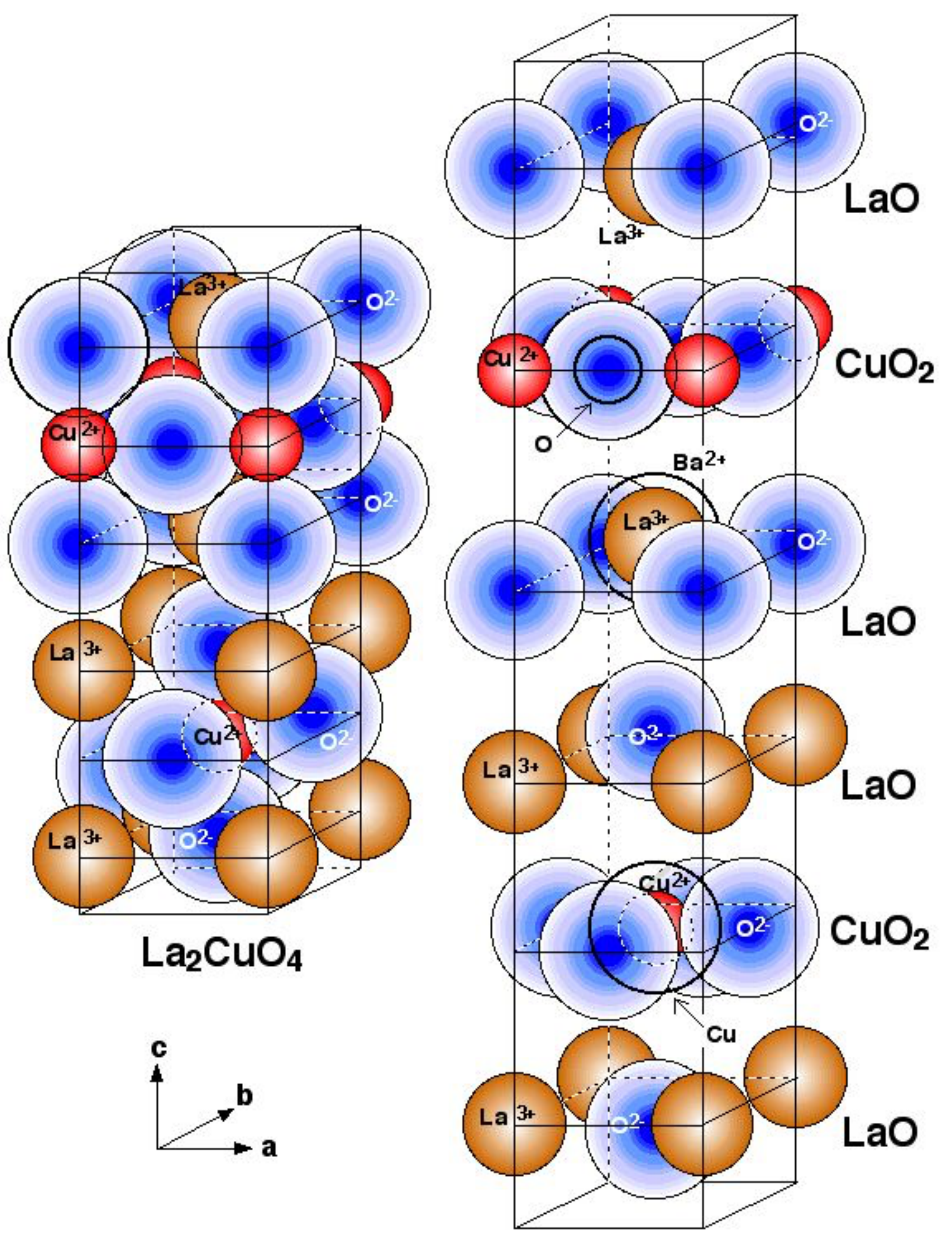} \footnotesize 

\noindent FIG. 1. Two unit cells of $La_2CuO_{4}$, staggered by half the planar lattice constants, ($\frac{a_0}{2},\frac{b_0}{2}$), in hard-sphere ion display (left), and vertically exploded (right) to better show the ion planes as noted on the side. When doped, $La^{3+}$ ions are substituted by larger-sized $Sr^{2+}$ or $Ba^{2+}$, as illustrated on the right (third plane from the top). Size and position of a lattice-defect $O$ atom that hosts two doped holes is also shown in the second plane. Likewise, a lattice-defect $Cu$ atom that hosts two doped electrons is shown in the second plane from the bottom. A $CuO_6$ octahedron can be seen in the bottom unit cell. (Figure 1 is placed here to facilitate a succinct reading of Sect. I.)
\normalsize 

\section{STRIPES IN $\mathbf{p}$-DOPED LANTHANUM CUPRATES }

The unit cell of pristine ${La_2CuO_4}$ has a central $CuO_2$ plane, sandwiched by $LaO$ layers (see Fig. 1). Consider the formation of the crystal by step-wise ionization, where brackets indicate electron localization at atoms, both within the planes and by transfer from the $LaO$ layers to the $CuO_2$ plane:
\bigskip 
\newline\noindent $LaO \;\;:\; La^{3+} + 3e^- + \;\;O \rightarrow  La^{3+} + [2e^- + O]\; + 
\downarrow \overline {e^-\;} | \rightarrow  La^{3+} + O^{2-}$
\newline
$CuO_2 :\; Cu^{2+} + 2e^- + 2O \rightarrow  Cu^{2+} + [2e^- + O] + \;\;\;O \;\;\;\rightarrow  O^{2-} \;+  Cu^{2+} + O^{2-}$
\newline
$LaO \;\;:\; La^{3+} + 3e^- + \;\;O \rightarrow  La^{3+} + [2e^- + O]\; + 
\uparrow \underline {e^-\;} | \rightarrow  La^{3+} + O^{2-}$
\bigskip

\noindent In the simplest case of doping with alkaline-earth, $Ae = Sr, Ba$, divalent $Ae$ substitutes, in some cells, trivalent $La$ in both sandwiching layers: 

\bigskip 
\noindent $AeO \;\;:\; Ae^{2+} + 2e^- + \;\;O \rightarrow  Ae^{2+} + [2e^- + O]\; \;\;\;\;\;\;\;\;\;\;\;\; \;\rightarrow  Ae^{2+} + O^{2-}$
\newline
$CuO_2 :\; Cu^{2+} + 2e^- + 2O \rightarrow  Cu^{2+} + [2e^- + O] + \;\;\;O \;\;\;\rightarrow  O^{2-} \;\;+  Cu^{2+} + \mathbf{\tilde{O}}$
\newline
$AeO \;\;:\; Ae^{2+} + 2e^- + \;\;O \rightarrow  Ae^{2+} + [2e^- + O]\; \;\;\;\;\;\;\;\;\;\;\;\; \;\rightarrow  Ae^{2+} + O^{2-}$
\bigskip

\noindent Compared to the $La_2CuO_4$ host, $Ae$-doping causes a lack of transferred electrons, of density $p=x$, from the sandwiching $LaO$ layers to the $CuO_2$ planes---also considered as ``hole doping.''
This leaves some oxygen in the $CuO_2$ planes as \emph{neutral} atoms (marked bold above and below). They can be regarded as housing (pairs of) the holes.
Up to a density $\check{p} \le 0.02$, such holes are \textit{itinerant}, enabling $\mathbf{\tilde{O}}$ atoms to skirmish long-range antiferromagnetism (3D-AFM) and cause its collapse at $\check{p}$. The remaining lack of electron transfer leaves more oxygen atoms \textit{stationary} at lattice sites, $\mathbf{O}$. They give rise to static stripes (explained instantly):

\bigskip 
\noindent $AeO \;\;:\; Ae^{2+} + 2e^- + \;\;O \rightarrow  Ae^{2+} + [2e^- + O]\; \;\;\;\;\;\;\;\;\;\;\;\; \;\rightarrow  Ae^{2+} + O^{2-}$
\newline
$CuO_2 :\; Cu^{2+} + 2e^- + 2O \rightarrow  Cu^{2+} + [2e^- + O] + \;\;\;O \;\;\;\rightarrow  O^{2-} \;\;+  Cu^{2+} + \mathbf{O}$
\newline
$AeO \;\;:\; Ae^{2+} + 2e^- + \;\;O \rightarrow  Ae^{2+} + [2e^- + O]\; \;\;\;\;\;\;\;\;\;\;\;\; \;\rightarrow  Ae^{2+} + O^{2-}$
\bigskip

\noindent Because of their double holes, the skirmishing and stationary oxygen atoms, $\mathbf{\tilde{O}}$ and $\mathbf{O}$, appear positive, \textit{relative} to the host crystal. Coulomb repulsion spreads the double holes (residing in the oxygen atoms) to form a planar superlattice of $\mathbf{O}$ crystal defects.
Its periodicity---incommensurate with the crystal lattice and therefore called the ``incommensurability''---is given, in reciprocal lattice units (r.l.u.), by
\pagebreak

\noindent  
\begin{equation}
q_{c,m}^p(x)\Big|_{CuO_2}= s_{c,m} \frac{\Omega^{\pm}}{{4}}\sqrt {x - \check{p}}\;,\;\;\;\;\;x \le \hat{x}  \; .
\end{equation}
\noindent 
The formula is valid for stripes in the $CuO_2$ planes, as indicated, and for doping up to a ``watershed'' concentration $\hat{x}$, which depends on the species of doping and co-doping. The suffix $c$ stands for charge order and $m$ for magnetization. The stripe-kind factor is $s_c=2$ or $s_m=1$ and the stripe-orientation factor is $\Omega^{+}=\sqrt{2}$ for $x > x_6 = 2/6^2  \simeq 0.056$ when stripes

\bigskip \medskip

\includegraphics[width=6.5in]{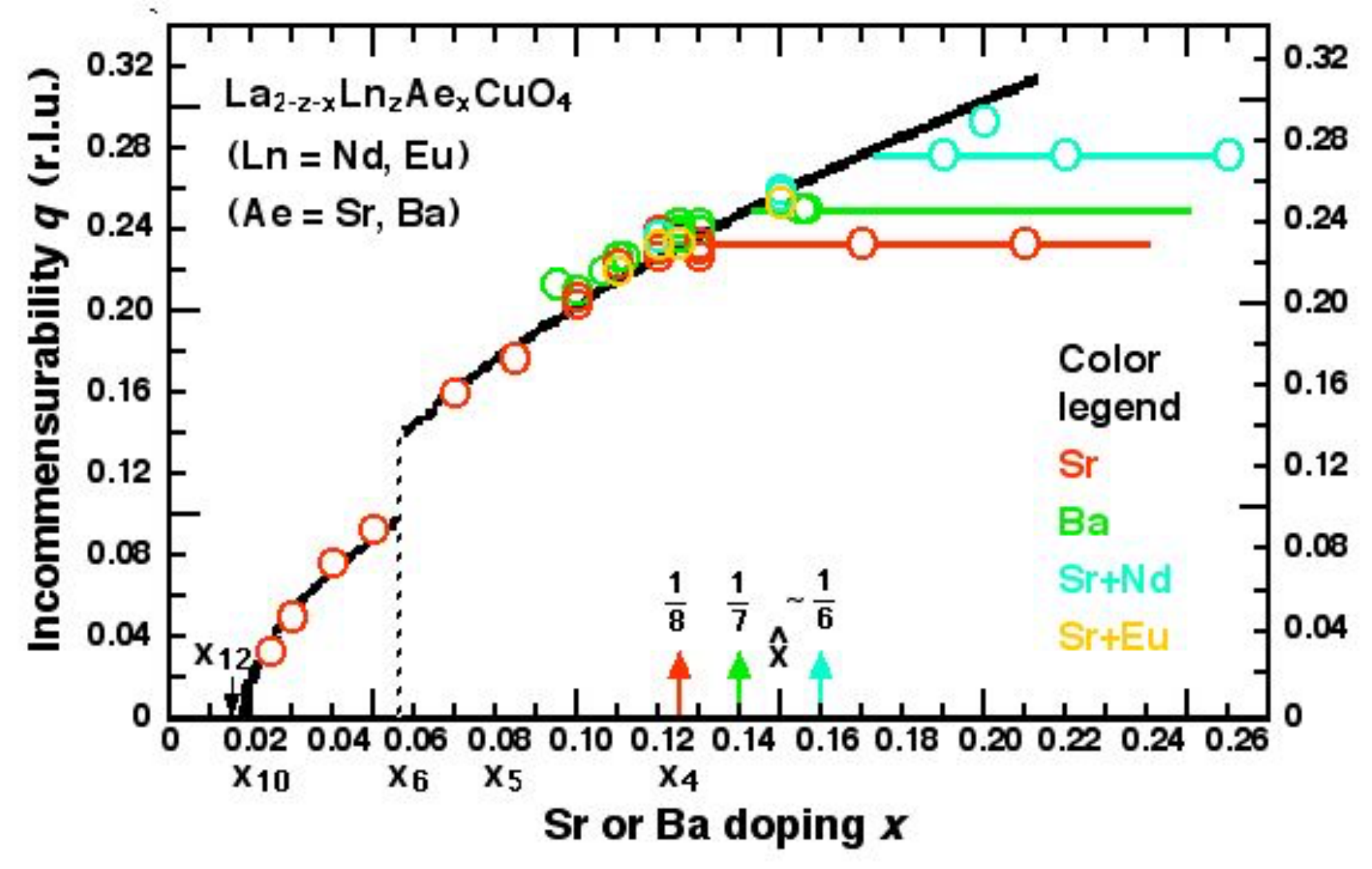}  \footnotesize 

\noindent FIG. 2. Incommensurability of charge-order stripes, $q = q_c$, and of magnetization stripes, $q = 2q_m$, in $La_{2-z-x}Ln_zAe_{x}CuO_{4}$ ($Ln = Nd, Eu; z = 0, 0.4, 0.2$) due to doping with $Ae = Sr$ or $Ba$. Circles show data from X-ray diffraction or neutron scattering (Refs. 7 - 48). 
The broken solid curve is a graph of Eq. (1), calculated with a constant offset value, $\check{p} = 0.02$. 
Commensurate doping concentrations are denoted by $x_n \equiv 2/n^2$.
The discontinuity at $x_6 \simeq 0.056$ is caused by a change of stripe orientation, relative to the planar crystal axes, from diagonal for $x<x_6$ to parallel for $x>x_6$. 
The curve holds for temperature at (and sufficiently near) $T=0$ and is accurate for low doping, $x<0.09$. Neglect of the doping dependence of the offset value, $\check{p}(x) < 0.02$, causes the slight deviation of the curve (too low) from most data in the doping range $x > 0.09$. 
Doping beyond watershed concentrations, $\hat{x}_{Sr} = 0.125 = 1/8$,  $\hat{x}_{Ba} = 0.14 \simeq 1/7$ and $\hat{x}_{Sr+Nd} = 0.17 \approx 1/6$, yields  constant stripe incommensurabilities,
$q_c(x) = 0.235$ ($Sr$), 0.25 ($Ba$) and 0.278 ($Sr$+$Nd$), respectively, given by Eq. (2) (dashed horizontal lines).
\normalsize 

\noindent   are parallel to the $a$ or $b$ axis, but $\Omega^{-} = 1$ for $x < x_6$ when stripes are diagonal. The offset value $\check{p}$ under the radical is the hole concentration necessary to keep 3D-AFM suppressed.
We may regard these holes as ``suppressor holes'' (mnemonically indicated by the over-dent of $\check{p}$). They reside in the itinerant $\tilde{O}$ atoms whose skirmisher task keeps the latter from participating in charge-order stripes. 

The derivation of Eq. (1) is based on a partition of the $CuO_2$ plane by \emph{pairs} of doped holes, incorporating the observed stripe orientation, in tetragonal approximation of the lattice constants, $a_0 = b_0$ (see Appendix A).\cite{1} The equation is valid at temperature $T \approx 0$. As no directional preference is used in the derivation, one would expect the corresponding charge-order and magnetization pattern to be checkerboard-like rather than stripe-like, as observed. The unidirectional character is imposed by the low-temperature phases of $La_{2-z-x}Ln_zAe_{x}CuO_4$ crystals. In these phases, $CuO_6$ octahedra are slightly tilted parallel or diagonal to the planar crystal axes to reduce stress from lattice mismatch due to ion-size differences ($Ba^{2+} >Sr^{2+} \approx La^{3+}$), with the  same tilt for whole crystal domains.\cite{5,6} This creates a preference of charge-order and magnetization pattern in one direction over the orthogonal one, resulting in unidirectional stripes. 

A large host of data\cite{7,8,9,10,11,12,13,14,15,16,17,18,19,20,21,22,23,24,25,26,27,28,29,30,31,32,33,34,35,36,37,38,39,40,41,42,43,44,45,46,47,48} from neutron scattering, hard X-ray diffraction, and resonant soft X-ray scattering is well described by Eq. (1) (see Fig. 2). 
For low temperatures and low doping ($x < 0.09$), the offset value $\check{p}$ in Eq. (1) agrees with the N\'{e}el concentration, $\check{p} = x_{N0}$, defined by vanishing N\'{e}el temperature, $T_N(x_{N0}) \equiv 0$. In $La_{2-z-x}Ln_zAe_{x}CuO_4$ compounds it has a value $x_{N0} = 0.02$, marking 
the collapse of 3D-AFM at $T=0$.
With more $Ae$ doping, but still at $T\approx0$, it is found that a \emph{smaller} value, $\check{p} < x_{N0}$, suffices to keep 3D-AFM suppressed.\cite{2}
Thus the use of $\check{p} = 0.02$ in Eq. (1) becomes inaccurate
beyond the low doping range, $x > 0.09$, as it gives \emph{too small} a value for the incommensurabilty $q_{c,m}(x)$. This can be seen in Fig. 2 where in that range most data points cluster slightly above the drawn $q(x)$ curve. 
Use of a \emph{diminished} offset value, $\check{p} \simeq 0.015$, in this range shifts that section of the curve slightly upward to better agreement with experiment (not shown). Specifically in the cases of the newly measured data\cite{9,10,11} from $La_{1.875}Eu_{0.2}Sr_{0.125}CuO_{4}$ and $La_{2-x}Ba_{x}CuO_{4}$ ($x = 0.115,\; 0.125$) at low temperature, 
the offset values are calculated as $\check{p} = 0.016, \; 0.014, \; 0.015$, respectively, instead of $\check{p} = 0.02$ (see Table II).
An explanation for the reduction of the offset value (at $T \approx 0$) with more doping $x$ will be given shortly. 

Historically, Eq. (1) was preceded by the empirical, ramp-like ``Yamada relation'' for magnetization stripes (also called spin density waves), inferred from neutron scattering experiments with $La_{2-x}Sr_xCuO_4$.\cite{3} 
It states that $q_m(x) = x$ for $x < 0.12$ but levels off to $q_m \approx 0.125=1/8$ for larger doping.
The magnetization stripes were joined by charge-order stripes (also called charge density waves) after their discovery in $La_{1.6-x}Nd_{0.4}Sr_xCuO_4$ by Tranquada \textit{et al.}\cite{4} 
The two types of stripes were observed to be related as $q_c(x) = 2 q_m(x)$.
In contrast to the multitude of experimental data for $x \le 0.125$, few data for larger doping had been available until recently, 
with some data falling close to the square-root curve but others considerably below. 
New data\cite{7,8,9,10,11,12,13,14,15} from resonant inelastic X-ray spectroscopy (RIXS) and neutron scattering on $La_{2-z-x}Ln_zAe_{x}CuO_{4}$ in the doping range $0.115 \le x \le 0.26$ have helped to clarify the situation, amounting to a qualitative change of the incommensurability at watershed doping concentrations $\hat{x}$ that depend on the doping (and co-doping) species. 
The qualitative change shows up as \emph{kinks} in the $q_{c,m}(x)$ profile at $\hat{x}$, where the square-root curve from Eq.(1) levels off to constant plateaus,
\begin{equation}
    q_c^p(x)\Big|_{CuO_2} = \frac{\sqrt{2}}{2} \sqrt{\hat{x} - \check{p}} \;,\;\;\;\; x > \hat{x} \;,
\end{equation}
with values $\hat{x}_{Sr} = 0.125 = 1/8$, $\hat{x}_{Ba} = 0.14 \simeq 1/7$, and $\hat{x}_{Sr+Nd} = 0.17\approx 1/6$. The corresponding constant incommensurabilities are $q_c(Sr) = 0.235$, $q_c(Ba) = 0.25$, and $q_c(Sr$+$Nd)=0.278$ (dashed lines in Fig. 2). 
Although the linear slope of Yamada's ramp only holds approximately, the prediction of a ramp \textit{level} has been confirmed.

The qualitative change of the incommensurability is caused by saturating Coulomb repulsion between the doped holes. Hole-doping of the $CuO_2$ plane builds up charge density, $\sigma = |e| x$, in domains of area $A$. The electrostatic work,
\begin{equation}
W(x) = \int^{A} da \int_0^{\sigma} \sigma' d\sigma' = C x^2 \;,
\end{equation}
 makes additional hole-doping of the $CuO_2$ planes progressively more costly. 
(The constant $C$ in Eq. (3) contains the elementary charge, $|e|$, and geometry factors.\cite{49})
Hole-doping of the $CuO_2$ plane ceases when its cost reaches the cost of doping holes into the $LaO$ layers, where they also reside pairwise in $O$ atoms.
This happens beyond the watershed doping $\hat{x}$.
Again, Coulomb repulsion spreads the double holes to a superlattice of lattice-defect $O$ atoms with attending charge order-stripes of incommensurability

\begin{equation}
    q_c^p(x)\Big|_{LaO}  = \frac{\sqrt{2}}{2}\sqrt {\frac{x - \hat{x}}{2} }\;,\;\;\;\;\;  x > \hat{x} \; .
\end{equation}
The denominator 2 under the radical accounts for the \emph{two} $LaO$ layers per unit cell.
No offset value $\check{p}$ occurs in Eq. (4), as suppresion of 3D-AFM is already accounted for by Eq. (2).
Strictly speaking, there is a slight increase of $q_c(x)$ in the $CuO_2$ plane for $x >\hat{x}$, over the value from Eq. (2) because hole-charge density builds up also in the $LaO$ layers. However, it is negligible because of both the $x^2$ dependence of the Coulomb term in the $CuO_2$ plane, Eq. (3), and the {\it two} $LaO$ layers per unit cell, Eq. (4).

For the samples that are doped with $Ae$ \emph{only}, a small difference of the watershed concentrations is observed: $\hat{x}_{Sr} = 0.125 = 1/8$ but $\hat{x}_{Ba} = 0.14 \simeq 1/7$. 
What could be the reason for the difference? Basically, the reason must originate with the different size (ionic radius) of host and doping cations, $r(La^{3+}) \approx r(Sr^{2+}) < r(Ba^{2+})$. The crystal relieves internal stress from ion-size mismatch by assuming various phases, depending on temperature and doping. The doping dependence of the low-temperature orthorhombic (LTO) and low-temperature tetragonal (LTT) phase, $0 \le$ LTO $<x_{LT}<$ LTT, is given by the phase boundary, which is at $x_{LT} = 0.21$ for $Sr$ doping, but in the range $0.11 <x_{LT} \le 0.125$ for $Ba$ doping (depending on $T$).\cite{5,6,46} The different phases in the doping range of interest---LTO of $La_{2-x}Sr_xCuO_4$ but LTT of $La_{2-x}Ba_xCuO_4$---may account for the different watershed values $\hat{x}_{Sr}$ and $\hat{x}_{Ba}$.
It is known that some properties of lanthanum cuprates extend to higher $Ae$ doping when co-doped with $Ln$. A well-known example is the (temperature dependent) doping level $x^*(T)$ where the pseudogap closes,
\begin{equation} 
x^*_{Sr+Ln}(T) = x^*_{Sr}(T) + \Delta x^*_{Ln} \;,\;\;\;\; Ln = Eu, Nd \;,
\end{equation}
which extends by $\Delta x^*_{Ln} \approx 0.05$ when co-doped.\cite{33,50} 
It reflects the extended incommensurability, Eq. (2), of co-doped samples,
$\hat{x}_{Sr+Nd} - \hat{x}_{Sr} = 0.170 - 0.125 = 0.045$. 

The close agreement of Eq. (1) with experiment lends credence to the underlying superlattice concept.
 The spacing of the planar square superlattice is reciprocal to the incommensurability of the charge density stripes, $L_c(x) = 1/q_c(x)$.
Together with Eq. (1) we obtain the density of the superlattice-forming holes,
\begin{equation}
p - \check{p} = \frac{2}{L_c(x)^2} \;,
\end{equation}
as a \emph{pair} of holes per superlattice unit square.
In order to localize on the superlattice, \emph{two holes} must reside together on a superlattice site. This leaves as the only viable choice that the defect superlattice is formed by \textit{neutral oxygen} atoms, $2e^+ + \; O^{2-}   \rightarrow  O$.

The presence of \textit{two} holes in $O$ atoms, as opposed to single-hole residence in $O^-$ ions, $e^+ + \; O^{2-} \rightarrow O^-$ (and as previously assumed\cite{1,2}), comes as a surprise. Apparently the pairing of holes in the $O$ atoms, combined with the Coulomb repulsion between the $O$ atoms (of charge $+2|e|$ relative to the host crystal), is energetically favorable over Coulomb repulsion between $O^-$ ions (of relative charge $+1|e|$) that would occur for a correspondingly denser $O^-$ distribution. The question of whether $O$ atoms or $O^-$ ions constitute the charge-order stripes in the $CuO_2$ planes, could have some bearing on superconductivity (e. g. pair-density waves). It is therefore desirable to corroborate (or refute) the double occupancy of doped holes in $O$ atoms with experiments other than stripe incommensurability. 

3D-AFM in the $La_2CuO_4$ host is borne by $\mathbf{m}(Cu^{2+})\ne0$ moments only, as both the $La^{3+}$ and $O^{2-}$ ions have a vanishing magnetic moment, $\mathbf{m}(La^{3+})=\mathbf{m}(O^{2-})=0$, due to their closed electron shells.
 In a free oxygen atom, $O$, the electrons in the $2p^4$ subshell arrange 
 their spin as $[\uparrow\downarrow]$ $[\uparrow]$ $[\uparrow]$, in agreement with Hund's rule of maximum multiplicity, rather than $[\uparrow\downarrow]$ $[\uparrow]$ $[\downarrow]$ or $[\uparrow\downarrow]$ $[\uparrow\downarrow]$ $[\;\;]$.
Thus the atom's spin quantum number is $S = 2\times \frac{1}{2} = 1$ (spin-triplet state). The same spin constellation can be assumed in lattice-defect oxygen atoms---be they itinerant, $\tilde{O}$, or stationary, $O$---which then have a non-vanishing magnetic moment, $\mathbf{m}(\tilde{O}) =\mathbf{m}(O) \ne 0$. 
The itinerant $\mathbf{m}(\tilde{O})$ moments skirmish the ordered $\mathbf{m}(Cu^{2+})$ moments of the host lattice and cause the collapse of 3D-AFM at $Ae$ doping $x_{N0}=x_{10} = 0.02$. However, 3D-AFM still needs to be kept suppressed at higher $Ae$ doping by a contingent of $\tilde{O}$ skirmishers with density $\check{p}$ of suppressor holes, Eq. (1).

The moments $\mathbf{m}(O)$ of neighboring $O$ atoms (with respect to the $O$ superlattice) arrange themselves antiparallel, which gives rise to their antiferromagnetic ordering (magnetization stripes) in the $CuO_2$ planes.
Because of this ordering, their spatial period is twice that of the charge-order stripes. Accordingly the incommensurabilities are $q_m(x) = \frac{1}{2} q_c(x)$, Eq. (1). This furnishes a natural explanation for the coupling of the incommensurabilities in the ratio of 1:2. In other words, the entities that form the corresponding stripes---be they electric charges, be they magnetic moments---reside at the \emph{same sites}, in this case at $O$ atoms, with the \emph{proviso} that the magnetic moments are in antiparallel order.

With increasing doping, the antiferromagnetic (afm) ordering of the $\mathbf{m}(O)$ moments of the magnetization stripes weakens the afm ordering of the
$\mathbf{m}(Cu^{2+})$ moments of the surviving 2D-AFM (spin glass, nematicity) of the host. 
A lower concentration of $\tilde{O}$ skirmishers
is then necessary to keep 3D-AFM suppressed.
This provides a likely explanation for the diminished offset value, $\check{p} \simeq 0.015$ for $x > 0.09$. 
At low doping, $x_{10} < x < x_6$, the $\mathbf{m}(Cu^{2+})$ moments of the 2D-AFM  in the $CuO_2$ planes continue to be diagonally orientated, with respect to the planar axes, like the 3D-AFM in pristine $La_{2}CuO_{4}$. They impose a diagonal orientation on the $\mathbf{m}(O)$ moments of the fledgling magnetization stripes. With closer proximity to one another at more doping, $x > x_6$, the $\mathbf{m}(O)$ moments break free to assume axes-parallel orientation of the magnetization stripes (causing the discontinuity in Fig. 2).

Although of no practical relevance, a qualification of Eq. (1) needs to be mentioned. At low $Ae$-doping $x < x_6$, only magnetization stripes, but
no charge-order stripes are observed, neither with hard X-ray diffraction, nor with the more sensitive method of resonant soft X-ray scattering (RXS).\cite{50} In contrast to dipoles---magnetic in magnetization stripes---point charges cannot be polarized. Thus charge order stripes from the $O$ superlattice are expected to be always axes-parallel oriented. Their non-observability in the $x < x_6$ doping range is likely caused by the diagonal orientation of the $\mathbf{m}(O)$ moments, upsetting the RXS scattering condition.

\section{STRIPES IN $\mathbf{n}$-DOPED LANTHANIDE CUPRATES }
For crystal formation of pristine $Nd_{2}CuO_{4}$, consider step-wise ionization, where the brackets indicate electron localization at atoms, both within the planes and by transfer from the $NdO$ layers to the $CuO_2$ plane:

\bigskip 
\noindent $NdO \;\;:\; Nd^{3+} + 3e^- + \;O \rightarrow  Nd^{3+} + [2e^- + O]\; + 
\downarrow \overline {e^-\;} | \rightarrow  Nd^{3+} + O^{2-}$
\newline
$CuO_2 :\; Cu^{2+} + 2e^- + 2O \rightarrow  Cu^{2+} + [2e^- + O] \;+ \;\;\;O \;\;\;\rightarrow \; O^{2-} \;\;+  Cu^{2+} + O^{2-}$
\newline
$NdO \;\;:\; Nd^{3+} + 3e^- + \;O \rightarrow  Nd^{3+} + [2e^- + O]\; + 
\uparrow \underline {e^-\;} | \rightarrow  Nd^{3+} + O^{2-}$
\bigskip

\noindent In the simplest case of {\it doping},
$Ce$ substitutes, in some cells, $Nd$ in both sandwiching layers: 

\bigskip 

\noindent $CeO \;\;:\; Ce^{4+} + 4e^- + \;\;O \rightarrow | \overline {e^-} \downarrow + \; Ce^{4+} + [2e^- + O]\; + 
\downarrow \overline {e^-\;} | \rightarrow  Ce^{4+} + O^{2-}$
\newline
$CuO_2 :\; Cu^{2+} + 2e^- + 2O \rightarrow  Cu^{2+} + 
\;\;\;\;\;\;\;\;\;\;\;\;[2e^- + O] + \;\;\;O \;\;\;\rightarrow    \mathbf{Cu} \;+ \;2O^{2-}$
\newline
$CeO \;\;:\; Ce^{4+} + 4e^- + \;\;O \rightarrow | \underline {e^-} \uparrow +\; Ce^{4+} + [2e^- + O]\; + 
\uparrow \underline {e^-\;} | \rightarrow  Ce^{4+} + O^{2-}$
\bigskip

\noindent The transfer of excess electrons from doped $Ce$ atoms to the $CuO_2$ plane---called ``electron doping''---reduces some $Cu^{2+}$ ions to neutral $Cu$ atoms (marked bold) that harbor \emph{pairs} of electrons.
Having an unpaired electron spin, free $Cu$ atoms have the same magnitude of electron-spin magnetic moment as free $Cu^{2+}$ ions, $|\mathbf{m}_s(Cu)| = |\mathbf{m}_s(Cu^{2+})| = g_s s \mu_B =  \mu_B$, with Land\'{e} factor $g_s=2$, spin quantum number $s=\frac{1}{2}$, and Bohr magneton $\mu_B$.
In a crystal, various influences modify the magnetic moment of a crystal ion from the free-ion value---experimental values of $|\mathbf{m}(Cu^{2+})|$ in $La_2CuO_{4}$ and $YBa_2Cu_3O_{6+y}$ vary somewhat, but $|\mathbf{m}(Cu^{2+})| \approx 0.6 \mu_B$ is regarded as most reliable.\cite{53,54,55,56,57,58} 
Because of their same spin component, the magnetic moments of {\it crystalline} $Cu^{2+}$ and $Cu$ can be assumed to be comparable, $|\mathbf{m}(Cu^{2+})| \approx |\mathbf{m}(Cu)|$.
Thus the lattice-defect $Cu$ atoms, generated by electron doping, continue the afm order of the host crystal (until terminated by other circumstances).
This qualitatively explains the much higher stability of 3D-AFM in the electron-doped $Ln_{2-x}Ce_xCuO_{4}$ compounds, with N{\'e}el concentration $x_{N0} = 0.134$---in stark contrast to $x_{N0} = 0.02$ in hole-doped $La_{2-x}Ae_{x}CuO_4$.

Although not directly affecting the electron-doping of the $CuO_2$ planes, a structural difference of the sandwiching $LnO$ layers of $Ln_2CuO_{4}$ compounds ($Ln$ = $Nd, Pr, Sm$) from the corresponding $LaO$ layers of $La_2CuO_4$ should be mentioned.
The latter compound has the T-type structure, shown in Fig. 1, where the $O^{2-}$ ions of the $LaO$ layers reside at positions above and beneath the $Cu^{2+}$ ions (called ``apical''). This is in contrast to the T'-type structure of the $Ln_2CuO_{4}$ compounds where the $O^{2-}$ ions of the $LnO$ layers reside at positions above and beneath the $O^{2-}$ ions of the $CuO_2$ planes.\cite{52} (When $La_2CuO_4$ is $n$-doped, a T $\rightarrow$ T' transition occurs in $La_{2-x}Ce_{x}CuO_4$ at $Ce$-doping $x=x_6$.)

Similar to the case of hole doping, we assume that {\it pairs} of doped electrons reside in $Cu$ atoms. Coulomb repulsion spreads the doped electrons in each $CuO_2$ plane of ${Ln_{2-x}Ce_{x}CuO_4}$ to an incommensurate $Cu$ superlattice. Comparison with experiment will show the validity of the assumption. By analogy with hole doping, Eq. (1), we can expect a square-root dependence of the incommensurability on $Ce$ doping.
However, two circumstances need to be considered. 
(i) Because of $|\mathbf{m}(Cu^{2+})| \approx |\mathbf{m}(Cu)|$,  
 doped electrons (residing pair-wise in $Cu$ atoms) don't frustrate 3D-AFM. Thus no part of them is exempt from superlattice formation in order to keep 3D-AFM suppressed. Accordingly, there is no doping offset akin to the concentration of suppressor holes, $\check{p}$, as in Eq. (1).
(ii) It is well-known that electron-doped copper oxides need to be grown, for reasons of stability, in an oxygen atmosphere.\cite{52} As-grown $Ln_{2-x}Ce_{x}CuO_{4+\delta}$ typically contain excess oxygen of $\delta \approx 0.03$ which has to be removed subsequently by sufficient annealing in an $Ar$ atmosphere.\cite{59} This raises the possibility that the samples under consideration may still contain residual excess oxygen. It is believed that excess oxygen resides interstitially in the $LnO$ layers\cite{52,59} where it ionizes, $O \rightarrow  O^{2-}$, by taking electrons from the $CuO_2$ planes.
This amounts, besides electron doping {\it via} $Ce$, to additional \textit{hole} doping, $p = 2\delta$, which neutralizes a fraction of the doped-electron density, $\Delta n = -2\delta$, below the concentration of $Ce$, $n < x$. The consequence is a doping offset from excess oxygen, affecting the incommensurability in electron-doped compounds,\cite{2}
\begin{equation}
q_c^n(x) = \frac{\sqrt{2}}{{2}}\sqrt{x - 2\delta}\;,\;\;\;\;\;x > x_6  \;.  
\end{equation}
Because of the $Cu$ atoms' participation in the 3D-AFM of the host crystal, \emph{no} magnetization stripes occur in electron doped cuprates. Figure 3 shows incommensurabilities of charge-order stripes in ${Nd_{2-x}Ce_{x}CuO_{4+\delta}}$ and ${La_{1.92}Ce_{0.08}CuO_4}$ parallel to the planar  axes, observed with resonant X-ray scattering. The data fall close to the dashed curve from  Eq. (7)  for the case without excess oxygen ($\delta=0$) \emph{only} when $Ce$ doping is small or large, whereas the data in the middle range of doping fall beneath the dashed curve by about 15\%. However, for the case with $\delta=0.01$ oxygenation, the data closely skirt the \emph{solid} curve. This indicates that residual excess oxygen may be in the samples.

The asymmetry of the phase diagrams of hole-doped and electron-doped `214' compounds 

\medskip 
\includegraphics[width=5.3in]{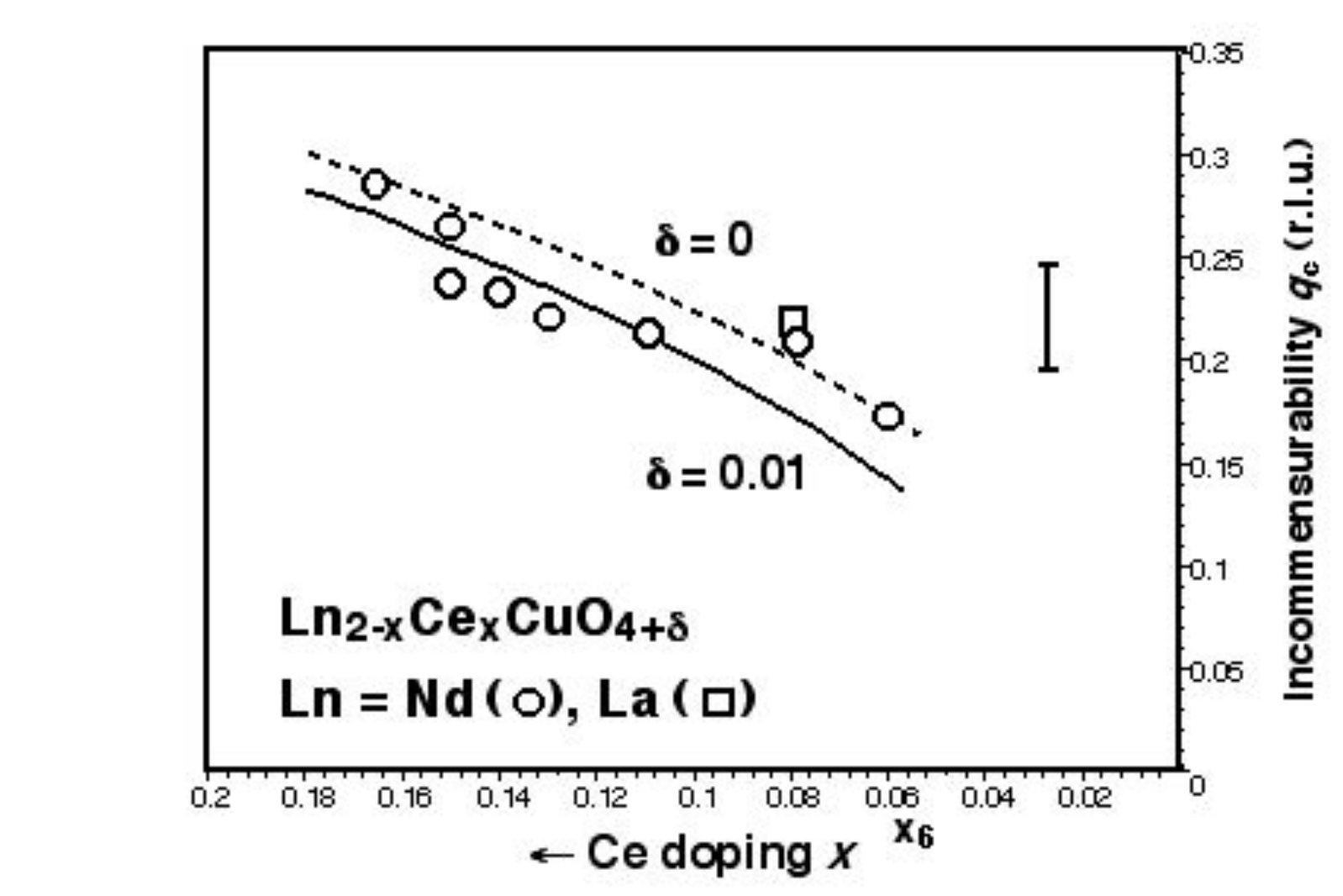}

\footnotesize
\noindent FIG. 3. Observed incommensurability $q_c$ of charge-order stipes in $Ln_{2-x}Ce_xCuO_{4+\delta}$ ($Ln = Nd$, crystal, circles; $Ln = La$, film, square) parallel to the planar axes (from Refs. 60 and 61). The curves are graphs of Eq. (7) without excess oxygen ($\delta=0$, dashed line) and with excess oxygen of $\delta=0.01$ (solid line). An average error bar of the the data is shown to the right. 
\normalsize 

\noindent has caused much discussion in the literature. Because of symmetry in doping---electron deficiency or excess---one would expect rather similar phase diagrams. In fact, one commonality---the square-root dependence of charge-order incommensurability  $q_c(x)$, Eqs. (1, 7)---results from the Coulomb repulsion of like charges. 
However, not only the \textit{charge} of doped holes or electrons is relevant, but also their \textit{residence}.
Many of the differences of the phase diagrams can qualitatively be explained by properties (size, magnetism) of the $O$ and $Cu$ atoms that house pairs of doped holes or electrons. 
It is conceivable that atomic size, $Cu > Cu^{2+},$ causes the T' instead of T lattice structure of $Nd_{2-x}Ce_xCuO_4$ at doping $x \ge x_6$, that is, when the superlattice spacing is $L(x) \le 6 a_0$.
To accommodate the large size of a lattice-defect $Cu$ atom (indicated in Fig. 1 in the bottom unit cell), $O^{2-}(2)$ ions above and beneath the $CuO_2$ plane would shift sideways from above/below $Cu^{2+}(1)$ ions to positions above/below $O^{2-}(1)$ ions.
As mentioned, similar magnetic moments, $|\mathbf{m}(Cu^{2+})| \approx |\mathbf{m}(Cu)|$, may account for the higher 3D-AFM stability as well as for the absence of magnetization stripes in the electron-doped compounds.
In contrast, the smaller size of $O$ than $O^{2-}$ makes \emph{no} T $\rightarrow$ T' transition necessary in the hole-doped compounds, and skirmishing $\mathbf{m}(\tilde{O})$ moments cause the rapid collapse of 3D-AFM.

As in the case of the hole-doped `214' compounds where an analysis of stripe incommensurability indicates the presence of hole {\it pairs} in lattice-defect $O$ atoms, it would be desirable to have independent experiments to check on the double occupancy of doped electrons in lattice-defect $Cu$ atoms in $Ln_{2-x}Ce_xCuO_4$ compounds. Such $Cu$ atoms should be observable by their hyperfine splitting, with microwave frequencies near the free-atom values, $\Delta \nu(^{63}Cu) = 11.734$ GHz and $\Delta \nu(^{65}Cu) = 12.569$ GHz.\cite{62}

\section{WATERSHED FOR STRIPES IN SUPERCONDUCTIVITY}
A controversy of long standing has been whether or not stripe order competes with superconductivity. When cooling $YBa_2Cu_3O_{6+y}$, the intensity of the X-ray signal from stripe order increases until the transition temperature to superconductivity, $T_c$, is reached, but then \emph{decreases} with more cooling (see Fig. 4).\cite{63,64} 
This ``cooling-curve break'' has been interpreted as a competition between stripe order and superconductivity. The experiment was soon repeated with $La_{2-x}Sr_xCuO_4$ samples by several research groups, yielding mixed results (see Table I). 
Recent experiments with $La_{2-x}Sr_xCuO_4$ show, upon cooling through $T_c$, continuing increase of the signal from charge-order stripes for $x = 0.115, \; 0.12, \; 0.13$ but a \emph{decrease} of the signal for $x = 0.144, \; 0.16$.\cite{12} 
The new findings present another watershed in the $Sr$ doping, occurring between $x = 0.13$ and $0.144$. 
We want to denote the watershed $Sr$ concentration of the cooling-curve break by $\hat{X}_{Sr}$.
It is close---but not equal---to the watershed of the compound's stripe incommensurability, $\hat{X}_{Sr} \approx \hat{x}_{Sr}=0.125$. 

A clear increase of the cooling curve below $T_c$ was also found for $La_{2-x}Ba_xCuO_4$ with $Ba$ doping $x = 0.095$ and 0.11 (see Table I).\cite{46}
The flat cooling curves for the doping values that flank the $Ba$ concentration $x = 0.125 = 1/8$---that is, $x = 0.115$ and 0.135---may be affected by the so-called ``1/8 anomaly'' where superconductivity is strongly suppressed, $T_c(\frac{1}{8}) \simeq 3$ K.
If the measurement of the $x=0.155$ doped crystal has validity (despite its low statistical significance), it would indicate a cooling-curve watershed in $La_{2-x}Ba_xCuO_4$ in the range $0.135 < \hat{X}_{Ba} < 0.155$, in good coincidence with the stripe watershed $\hat{x}_{Ba}=0.14$. It seems therefore warranted to explore whether a common cause exists, and more specifically, whether $\hat{X} = \hat{x}$?

It is observed that the cooling does \emph{not} affect the incommensurability of the stripes. By Eqs. (1, 2) this means that the cooling does not affect the density $x-\check{p}$ or $\hat{x}-\check{p}$
of stripe-forming holes in the $CuO_2$ planes. The change of X-ray intensity must therefore be due to fluctuations of the $O$ superlattice that degrade the diffraction in terms of lesser correlation lengths. Instead of cooling down from $T_c \rightarrow 0$, as in experiment, the cooling curves are easier understood for the reverse process of \emph{warming} up from $T=0\rightarrow T_c$. In this case, increasing thermal fluctuations with rising temperature lead to steadily decreasing signal strength, with no break when passing through $T_c$. Turning to the other case, an increasing warming curve---or decreasing cooling curve---is, besides $T<T_c$, subject to the condition

\includegraphics[width=5.5in]{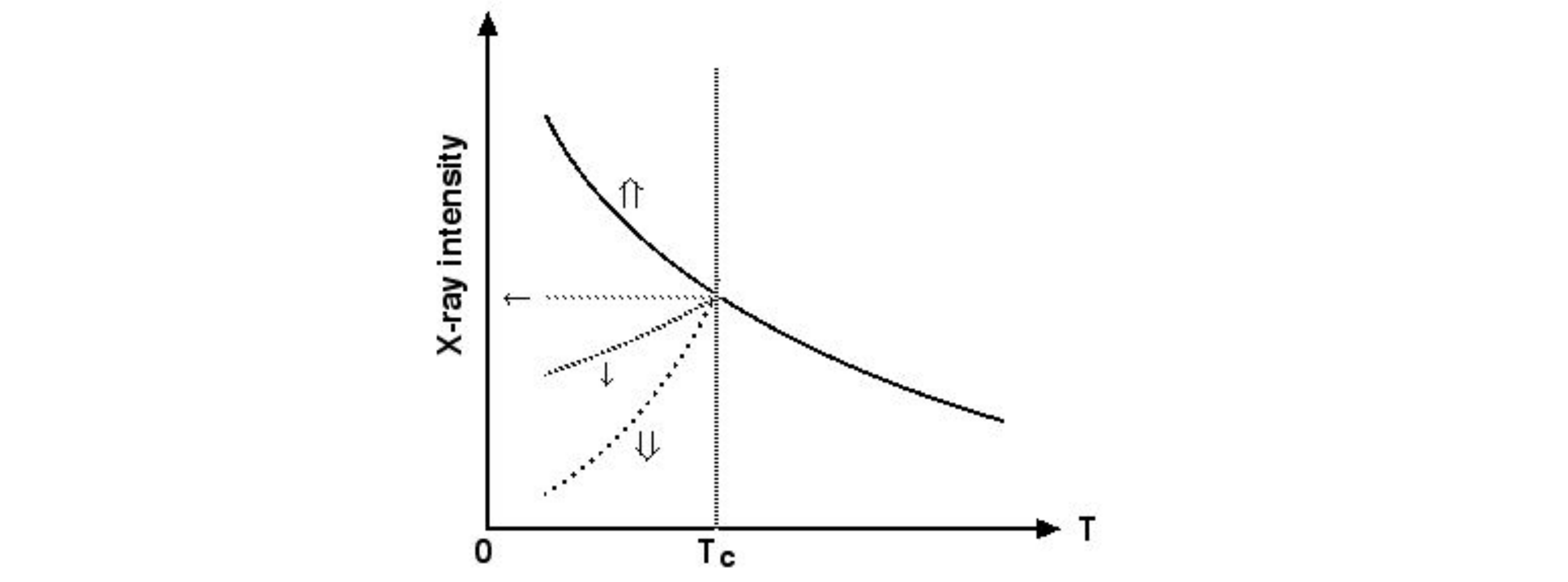}  \footnotesize

\noindent FIG. 4. Cooling curves (schematically) of X-ray intensity diffracted by charge-order stripes in the $CuO_2$ planes of
$La_{2-x}Ae_{x}CuO_{4}$ ($Ae=Sr, Ba$). The arrows correspond with Table I.
\normalsize 

\noindent $x > \hat{X}_{Ae}$. If the curve-break is caused by the stripe watershed, $\hat{X}_{Ae} = \hat{x}_{Ae}$, then the switch of double-hole residence from $O$ atoms in the $CuO_2$ plane to $O$ atoms in the $LaO$ layer becomes relevant.

In order to deal with the superconducting phase, we need to make assumptions about superconductivity in $La_{2-x}Ae_xCuO_4$. As a model, we want to assume that the superconducting state makes possible a scatter-free transfer of double holes back and forth between lattice defect $O$ atoms and their immediate $O^{2-}$ neighbor ions. At $Ae$-doping below the watershed, $x<\hat{x}$, the $O$  atom and the closest $O^{2-}$ neighbors assessable to the double holes are in the $CuO_2$ plane, located at face positions of the unit-square (see Fig. 1). Coulomb repulsion keeps the double-hole bearing $O$ atoms at their superlattice sites (on average when superconducting fluctuations set in). With increasing temperature, thermal agitation of lattice ions gives rise to larger fluctuations that degrade X-ray diffraction, diminishing their intensity. The result is a steadily decreasing warming curve---or increasing cooling curve---unimpeded by passing $T_c$.

With $Ae$ doping at and beyond the watershed, $x \ge \hat{x}$, the chemical potential of $O$ atoms in the $CuO_2$ plane and in the $LaO$ layers is equal. Now superconducting double holes can also fluctuate between an $O$ atom at a superlattice site in the $CuO_2$ plane and either one of the four $O^{2-}$ neighbors, located in the sandwiching $LaO$ layers as well as on the same unit-cell face (see Fig. 1). 
For example, if the $O$ atom is at a superlattice site ($\frac{1}{2},0,\frac{1}{2}$) in the $CuO_2$ plane, then such $O^{2-}$ ions are at positions ($\frac{1}{2}\pm\frac{1}{2},0,\frac{1}{2}\pm\frac{1}{2}$) in the $LaO$ layers. Fluctuations of that kind would considerably reduce the average presence of $O$ atoms in the $CuO_2$ planes and thus the intensity of the diffracted X-rays from the $CuO_2$ planes. 
The fluctuating presence of $O$ atoms in the $CuO_2$ planes does not change the stripe incommensurability $q_c$ because the superlattice spacing persists---the close fluctuation displacements to the sandwiching $LaO$ layers are negligible with respect to the superlattice spacing. The described effect is largest at $T=0$ but decreases to termination with $T \rightarrow T_c$ because of diminishing super/normal-conducting phase ratio (Gorter-Casimir two-fluid model of superconductivity).
The result is a steadily increasing warming curve up to $T_c$---or decreasing cooling curve down from $T_c$---where it joins the curve that is affected only by thermal fluctuations.

As Table I shows, the $q_c(x)$ values measured by different researchers are close, but minor differences persist. This can be seen best in a comparison of the data for $x=0.12$ (bold print). The disagreements may be due to differences in crystal growth, measurement equipment, data processing or background subtraction. Uncertainties are amplified when it comes to the trend of the cooling curves, and thus to the watershed doping $\hat{X}$. This may explain the different trends near $Sr$ doping $x=0.125$ as observed by different researchers. It is 
\bigskip

\begin{table}[ht]
\begin{tabular}{|p{1cm}|p{1.5cm}|p{1.5cm}|p{1.9cm}|p{1cm}|p{1cm}|p{4.6cm}|}
 \hline  \hline
Dop- ant &$\;\;\;x \;\;\;\;\;$ nominal &$\;\;\;q_c \;\;\;\;$   [r.l.u.] &  cooling curve $< T_c$ & Ref.&Year&Comment\\
 \hline  \hline
   Sr& $\mathbf{0.12}$  & 0.24* &$\;\;\;\;\;\;\;\Uparrow$   &66&2012&*only surface stripes \\
  \hline
Sr& 0.110  & 0.224 &$\;\;\;\;\;\;\;\Downarrow$   &33&2014& \\
Sr& $\mathbf{0.120}$  & 0.235 &$\;\;\;\;\;\;\;\Downarrow$   &33&2014& \\
Sr& 0.130  & 0.232 &$\;\;\;\;\;\;\;\Downarrow$   &33 &2014&\\
 \hline
Sr& $\mathbf{0.12}$  & $0.236^{\dagger}$ &$\;\;\;\;\;\;\;\downarrow$   &23&2014&$^{\dagger}$rotated 2.7$^\circ$ in the plane \\
 \hline
Sr&   $\mathbf{0.12}$  & 0.231 &$\;\;\;\;\;\;\;\leftarrow$   &24&2014& \\
 \hline
Sr&   $\mathbf{0.12}$  & $0.24(6)^{\dagger \ddagger}$ &not probed   &$\;\;8$&2017&$^{\ddagger}$ included for $q_c$ comparison \\
 \hline
Sr&  0.115  & $0.226$ &$\;\;\;\;\;\;\;\Uparrow$    &12&2019& \\
Sr& $\mathbf{0.12}$  & 0.2264 &$\;\;\;\;\;\;\;\Uparrow$   &12&2019& \\
Sr& 0.13 & 0.2281 &$\;\;\;\;\;\;\;\Uparrow$       &12&2019& \\
Sr& 0.144  & 0.2321 &$\;\;\;\;\;\;\;\Downarrow$   &12&2019& \\
Sr& 0.16  & 0.2322 &$\;\;\;\;\;\;\;\Downarrow$   &12&2019& \\
 \hline
 Sr& $\mathbf{0.12}$  & $0.232^{\ddagger}$ &not probed   &14&2020&$^{\ddagger}$ included for $q_c$ comparison \\
 \hline
Ba& 0.095  & 0.205 &$\;\;\;\;\;\;\;\Uparrow$   &46&2011& \\
Ba& 0.110  & 0.219 &$\;\;\;\;\;\;\;\Uparrow$   &46&2011& \\
Ba& 0.115  & 0.228 &$\;\;\;\;\;\;\;\leftarrow$   &46&2011& \\
Ba& $\mathit{0.125}$  & 0.232 &  $\;\;\;\;\; > T_c$ &46&2011& \\
Ba& 0.135  & 0.243 &$\;\;\;\;\;\;\;\leftarrow$   &46&2011& \\
Ba& 0.155  & 0.245 &$\;\;\;\;\;\;\;\;\downarrow ^\S$   &46&2011&$^\S$low statistics \\
 \hline \hline
\end{tabular}
\caption{Trend of the cooling curves of the X-ray intensity diffracted by charge-order stripes with incommensurability $q_c(x)$ in
$La_{2-x}Ae_{x}CuO_{4}$ ($Ae=Sr, Ba$) below the transition temperature $T_c$ (except for $La_{1.875}Ba_{0.125}CuO_{4}$). Double up-arrows and down-arrows signify clear increase and decrease, respectively. Single down-arrows signify moderate decrease and left-arrows indicate flat cooling curves. Cases with $x=0.12$ are marked bold for ease of comparison.}
\label{table:1}
\end{table}

\noindent therefore valuable to have cooling curves for $Ae$-doping $x$ distinctly away from the watershed $\hat{X}$. The series of cooling curves measured by Wen \textit{et al.}\cite{12} for $0.115 \le x \le 0.16$ shows a qualitative change in the curve profile with $Sr$ doping. Leaving  the measurement for $x=0.13$ temporarily aside (because of a re-calibration), the watershed value must lie in the doping interval $0.12 < \hat{X}_{Sr} < 0.144$.
When the $La_{2-x}Sr_xCuO_4$ crystal with nominal doping $x=0.13$ is re-calibrated (see Appendix B), an effective doping value $x^{eff}_{Sr}=0.121 < 0.125$ is obtained. 
This opens the possibility of a common origin for the watersheds in stripe pattern and cooling curves, $\hat{X}_{Sr} \simeq \hat{x}_{Sr} = 1/8 $.

What can be learned about stripe-superconductivity competition? Such competition was inferred from \textit{decreasing} cooling curves.
There seems to be now sufficient evidence, that \emph{no} decreasing cooling curves are observed for $Ae$ doping $x < \hat{X} \simeq \hat{x}$. Their observation for $x > \hat{X} \simeq \hat{x}$ may be attributed to superconducting fluctuations of double holes in $O$ atoms of both the $CuO_2$ planes and $LaO$ layers, as the present study proposes.

Another indication for stripe-superconductivity competition has been seen in the 1/8 anomaly of $La_{1.85}Ba_{0.125}CuO_4$, that is, the sudden drop of $T_c$ to 3 K. The attribution to charge-order stripes has been rationalized by the maximum of the charge-order and magnetization stripe observation temperature, $T_{cdw}(x)$ and $T_{sdw}(x)$, near $x=0.125$.\cite{46} However, it should be pointed out that the incommensurability of the stripes \emph{at} $x=0.125$, $q_c(0.125|Ba) \simeq 0.235$ and $q_m(0.125|Ba) \simeq 0.117$, is \emph{not} 1:4 or 1:8 commensurate with the crystal. It is not obvious why these incommensurate stripes would suppress superconductivity at $x=0.125$. 

More persuasive is an indication for stripe-superconductivity competition
in the $Sr$-doped compound, $La_{2-x}Sr_xCuO_4$, where the $T_{cdw}(x)$ and $T_{sdw}(x)$ domes,\cite{12,33,67} centered near $x=0.125$, coincide with the doping range of a big dent in the otherwise parabolic $T_c(x)$ dome. But $x<$ 0.125 is the doping range with rising cooling curves, which show $\emph{no}$ competition---the dent must have a different origin.

Another argument is based on an enhancement of diffracted X-ray intensity from stripes when superconductivity is destroyed with a strong magnetic field, first observed in $YBa_2Cu_3O_{6+x}$ ($x = 0.6, 0.67$).\cite{63,64} It is interpreted as ``suppressed suppression'' of stripes by superconductivity. The problem with this reasoning is that the introduction of another variable---the magnetic field---prevents a mono-causal conclusion.
The experiment was repeated for the same series of $La_{2-x}Ba_xCuO_4$ crystals where previously cooling curves without a magnetic field had been observed.\cite{65} The findings show that field enhancement is {\it relatively} large at $Ba$ doping $x = 0.125 \pm 0.03$, that is, where the stripe intensity is weak, but much smaller near $x = 0.125$ where stripe intensity is strong. The field effect is most pronounced at very low temperature ($T=3$ K) but fades rapidly with increasing temperature---much faster than the thermal decrease of stripe intensity without a field. 
A possible explanation of the findings could be that the field enhancement is caused by a slight upgrade of the diffraction condition of the $O$ superlattice, stiffened through stabilization of its magnetic moments $\textbf{m}(O)$ by a gain of Zeeman energy in the field.\cite{65} It is a small energy gain, essentially independent of doping $x$, but increasing with field strength, and
quickly overcome by thermal agitation with rising temperature. A view at the experimental spectra\cite{65} indicates that the {\it absolute} field enhancement is comparable
in the samples of the probed doping range, $0.095 \le x \le 0.155$, which would corroborate this interpretation.
Rather than compete with stripes, superconductivity passively stands by as thermal fluctuations eat away at stripe diffractibility.
Taken together, the counter-arguments leave little support for the notion of stripe-superconductivity competition in $La_{2-x}Ae_{x}CuO_{4}$---the case of $YBa_2Cu_3O_{6+y}$ is different.\cite{68}

\section{TEMPERATURE DEPENDENCE OF STRIPES}

Three of the recent articles address the temperature dependence of charge-order stripes in  $La_{2-x}Ba_{x}CuO_{4}$ ($x = 0.115, 0.125, 1.555$) and $La_{1.675}Eu_{0.2}Sr_{0.125}CuO_{4}$.\cite{9,10,11} 
Three aspects among the new experimental findings are:
\newline \#1 The incommensurability of charge-order stripes is essentially constant up to a threshold temperature $T'$, 
\begin{equation}
q_c(x,T \le T') \simeq q_c(x|T=0)\;, 
\end{equation}
\noindent as given by Eqs. (1, 2), but then increases linearly at higher temperatures,
\begin{equation}
q_c(x,T>T') = q_c(x,T\le T') + c \times (T-T')\;, 
\end{equation}
\noindent with a compound specific thermal coefficient $c > 0$ (depending slightly on doping $x$).\cite{9,10,11} 
\medskip

\noindent \#2 An exception is seen in the strontium-doped cuprate,
$La_{1.675}Eu_{0.2}Sr_{0.125}CuO_{4}$, below the threshold temperature, $T < T'$, where increasing temperature causes a slight \emph{decrease} of the incommensurability.\cite{11}

\noindent \#3 In the barium-doped cuprate $La_{1.875}Ba_{0.125}CuO_{4}$, the rigid coupling of the incommensurability of charge order stripes and magnetization stripes, $q_c(x)=2q_m(x)$, Eq. (1), ceases above the threshold temperature, $T > T'$. Whereas the incommensurability of the charge order stripes keeps increasing with temperature, Eq. (9), the double incommensurability of the \emph{magnetization} stripes \emph{decreases} by about the same rate,\cite{9} 
\begin{equation}
2q_m(x,T>T') = 2q_m(x,T\le T') - c \times (T-T')\;. 
\end{equation}

\section{EXPLANATION OF EXPERIMENTAL FINDING \#2}

In the attempt to explain these experimental findings we start with the second aspect---the slight \emph{decrease} of $q_c(T)$ with increasing temperature in $La_{1.675}Eu_{0.2}Sr_{0.125}CuO_{4}$ below the threshold temperature, $T < T'$. This appears as an exception to the general trend of constant incommensurability below $T'$, but of increasing values above, Eqs. (8, 9). The key of this ``exception'' lies in the offset hole concentration $\check{p}$ which appears under the radical in Eq. (1). 
At elevated temperature, less mobility of the skirmishing $\tilde{O}$ ions makes 3D-AFM suppression less effective, such that a diminished offset, $\check{p} \simeq 0.015$, is insufficent. Instead, the offset must approach the full value of the \emph{temperature-dependent} N\'{e}el concentration, $\check{p} \rightarrow x_N(T)$.
This can be seen in the case of $La_{1.675}Eu_{0.2}Sr_{0.125}CuO_{4}$, listed at the top of Table II: At low temperature, $T=25$ K, the offset value is clearly less than the N\'{e}el concentration, $\check{p} = 0.016 < 0.020 = x_{N0}$, but at elevated temperature, $T'=80$ K, the offset is calculated to be very close to the N\'{e}el concentration,
$\check{p}(80$K) = $0.020 \simeq x_N(80$K) $=0.019$.
[The comparison here is with $x_N(T')$ of $La_{2-x}Ae_xCuO_4$---co-doping with $Eu$ could slightly change its value.]
The approach of $\check{p} = 0.016 \rightarrow 0.020$ with increasing temperature below $T'$ is reflected in a slightly downward parabolic arc of the $q_c(T)$ display for $T = 25$ K$ \rightarrow$ 
\linebreak 80 K $=T'$ (Fig. 3a in Ref. 11), caused by the square-root dependence in Eq. (1).

\section{EXPLANATION OF EXPERIMENTAL FINDING \#1}

If the charge-order stripes are caused by the doped holes and if Eq. (1) adequately gives their incommensurability $q_c(x)$, then the observed $q_c$ value, together with the temperature-dependent N\'{e}el concentration for the offset value $\check{p} = x_N(T)$, can be used to calculate the hole density $p(T)$ at elevated temperatures. For $La_{1.675}Eu_{0.2}Sr_{0.125}CuO_{4}$ at $T = 210$ K this  gives a hole concentration of $p(T_{210K})= 0.152 > x = 0.125$, considerably larger than the $Sr$ doping (see Table II). Similarly, for $La_{2-x}Ba_{x}CuO_{4}$ ($x = 0.115,\; 0.125$) at $T = 49$ K and  
90 K, this gives hole concentrations of $p(T_{49K}) = 0.130 > 0.115$ and $p(T_{90K}) = 0.167 > 0.125$.

 Where do the additional holes come from? We may regard them as ``thermally generated holes.'' More specifically, they can be regarded as the positive electric  partners of thermally 
 generated electron-hole \emph{pairs}, required by charge neutrality of the crystal. A possible scenario could be a thermally activated transfer of an electron pair (here symbolized by $\rightarrow 2e^-$$\rightarrow$) from crystal $O^{2-}$ ions to neighbor crystal $Cu^{2+}$ ions,  
\begin{equation}
O^{2-} \;\; \rightarrow2\; e^-\rightarrow \;\; \;Cu^{2+} \;\;=\;\;\; O \;+\; Cu \;,
\end{equation}
leaving $O$ and $Cu$ atoms behind as lattice defects. The $Cu$ atoms then harbor the thermally generated electrons of concentration 
\begin{equation}
n^{\dagger}(T) = p^{\dagger}(T), \;\;\;\;\;\;\;\;\  T > T' \;.
\end{equation}
Adding the thermally generated holes to the holes introduced by $Ae$ doping yields the total hole concentration, 
\begin{equation}
p(T) = x + p^{\dagger}(T).
\end{equation}
 It gives rise to the observed charge order stripes of incommensurability,
\begin{equation}
q_c(x,T>T')\Big|_{CuO_2}  = \frac{\sqrt{2}}{2}\sqrt {p(T) - \check{p}}
\approx q_c(x,T\le T') + c \times (T-T')\; ,
\end{equation}
with $c= \frac{dp^{\dagger}}{dT}
\Big|_{T'}\large/ q_c(x,T\le T')$, 
obtained by Taylor expansion.

What happens to the thermally generated \emph{electrons}? Hosted pairwise by $Cu$ atoms (of charge $-2|e|$ relative to the host crystal), we assume
that they spread out to form a lattice-defect $Cu$ superlattice and corresponding charge order stripes with incommensurability
\begin{equation}
q_c(n^{\dagger})\Big|_{CuO_2}  = \frac{\sqrt{2}}{2}\sqrt {n^{\dagger}} \;  .
\end{equation} 
No setoff $\check{p}$ appears under the radical in Eq. (15)---no 3D-AFM suppression by $\mathbf{m}(Cu) \simeq \mathbf{m}(Cu^{2+})$. 
Because of their weakness and being embedded in thermal noise, the charge-order stripes from the thermally generated $Cu$ superlattice may be beyond detectibility.
No accompanying magnetization stripes occur. Instead, the $Cu$ atoms adversely affect the magnetization stripes from the $O$ atoms, as explained next.

 \section{EXPLANATION OF EXPERIMENTAL FINDING \#3}

Above the threshold temperature $T'$, the thermally generated electron-hole pairs separate and reside pairwise at $Cu$ and $O$ lattice defects, respectively, Eq. (11). 
Accordingly, more $\mathbf{m}(O)$ moments are \emph{thermally} generated, but also the same amount of $\mathbf{m}(Cu)$ moments, Eq. (12). 
As mentioned in Sect. II, $\mathbf{m}(Cu)$ moments in electron-doped $Ln_{2-x}Ce_xCuO_4$ compounds tend to align with the AFM of the $\mathbf{m}(Cu^{2+})$ moments of the host. In this vein, the \emph{thermally} generated $\mathbf{m}(Cu)$ moments in hole-doped $La_{2-x}Ae_xCuO_4$ tend to strengthen 2D-AFM, opposing the weakening effect from $\mathbf{m}(O)$ moments.
It depends on the relative magnitude of the $\mathbf{m}(O)$ and $\mathbf{m}(Cu)$ moments, 
\begin{equation}
r = \frac{|\mathbf{m}(Cu)|}{|\mathbf{m}(O)|}  \;,
\end{equation}
whether at $T > T'$ there is a \emph{net} increase or decrease of uncompensated $\mathbf{m}(O)$ moments, and accordingly of the incommensurability $q_m(x,T>T')$ of the magnetization stripes.

For example, if we had $r=1$, then each $\mathbf{m}(Cu)$ moment would compensate one antiparallel $\mathbf{m}(O)$ moment.
In general, each $\mathbf{m}(Cu)$ moment compensates (on average) $r\;\mathbf{m}(O)$ 

\bigskip 
 
 \begin{table}[ht!]

\begin{tabular}{|p{0.9cm}|p{1.0cm}|p{4cm}|p{1.2cm}|p{1.4cm}|p{1.cm}|p{1.5cm}|p{1.2cm}|p{1.65cm}|p{0.8cm}|  }
 \hline  \hline
$T$ [K]  & $T'$ [K] & $\;\;\;\;\;\;\;$Compound & $\;\;\;\;x $ & $\;\;\;p(T)$ & $\;\;\Delta p$ & $q_c$ [r.l.u.]  & $\; x_N(T)$ & $\;\;\;\;\;\;\check{p}$ & Ref.\\
 \hline  \hline
$\;25$  & 80  &$La_{1.8-x}Eu_{0.2}Sr_xCuO_{4}$ & 0.125 & =0.125 & &  0.233 & 0.020 & $\;\;\mathbf{0.016}$    &11 \\
$\;80$  & 80  &$La_{1.8-x}Eu_{0.2}Sr_xCuO_{4}$ & 0.125 &=0.125 & & 0.229 & 0.019 & $\;\;\mathbf{0.020}$   &11 \\
210  & 80  &$La_{1.8-x}Eu_{0.2}Sr_xCuO_{4}$ & 0.125 &$\;\;\; \mathbf{0.152}$ & 0.027 &  0.27 & 0.006 & =0.006    &11\\
 \hline
$\;20$  & 33  &$La_{2-x}Ba_{x}CuO_{4} $ & 0.115 & =0.115 & &  0.225 & 0.020 & $\;\;\mathbf{0.014}$   &10 \\
$\;33$  & 33  &$La_{2-x}Ba_{x}CuO_{4} $ & 0.115 & =0.115 & &  0.225 & 0.020 & $\;\;\mathbf{0.014}$  &10 \\ 
$\;49$  & 33  &$La_{2-x}Ba_{x}CuO_{4}$ & 0.115 &$\;\;\; \mathbf{0.130}$ & 0.015 &  0.240 & 0.020 & $\;\;\;0.015^*$    &10 \\
 & & & & & & & &*estimate  &\\
$\;23$  & 54  &$La_{2-x}Ba_{x}CuO_{4}$ & 0.125 & =0.125 & &  0.235 & 0.020& $\;\;\mathbf{0.015}$  & 9,10 \\
$\;54$  & 54  &$La_{2-x}Ba_{x}CuO_{4}$ & 0.125 & =0.125 & &  0.235 & 0.020 & $\;\;\mathbf{0.015}$   &9,10 \\
$\;90$  & 54  &$La_{2-x}Ba_{x}CuO_{4}$ & 0.125 &$\;\;\; \mathbf{0.167}$ & 0.042 &  0.272 & 0.019 & =0.019   &9,10\\

 \hline   \hline
\end{tabular}
\caption{Incommensurability $q_c$ of charge-order stripes in lanthanide cuprates of nominal doping $x$, measured with RIXS at temperature $T$.
The threshold temperature $T'$ denotes the onset of temperature-dependent increase of $q_c$. Experimental values of the N\'{e}el concentration $x_N(T)$ are from Ref. 69. The hole concentration $p(T)$ and the offset value $\check{p}$ are assumed to be equal to the preceding entry, indicated by the '=' sign, but otherwise \textbf{calculated} (bold print) with Eq. (13). The values in the $\Delta p$ column give the concentration of thermally generated holes, $\Delta p = p^{\dagger}(T)$.}
\label{table:2} \end{table}

\noindent  moments.
By Eqs. (1, 15, 16) the remaining density of uncompensated $\mathbf{m}(O)$ moments,
\begin{equation}
[\mathbf{m}(O)] \equiv \frac{x-\check{p}}{2} + \frac{p^{\dagger}}{2} - \frac{n^{\dagger}}{2} =\frac{x-\check{p}}{2} +  \frac{1-r}{2}p^{\dagger},
\end{equation} gives rise to a magnetic stripe incommensurability
\begin{equation}
2q_m(x,T > T') = 2\;\frac{\Omega^{\pm}}{4}\sqrt {x - \check{p} + (1-r)p^{\dagger}}  \approx 2q_m(x,T\le T') + (1-r) c \times (T-T').
\end{equation}
Comparison with the temperature dependence of charge-order stripes, Eq. (14), shows that for $r=2$ the double incommensurability of magnetic stripes would \emph{decrease} with increasing temperature by a rate comparable with the increase of the incommensurability of the charge density stripes. Qualitatively, this breaks the locking of charge order and magnetization stripes, $q_m(x) \ne \frac{1}{2}q_c(x)$, above the temperature threshold, $T > T'$.
Quantitatively, with $\mathbf{m}(Cu) \approx \mathbf{m}(Cu^{2+})$, 
what is missing for a validation of the present scenario of temperature-dependent magnetization stripes, Eq. (18), is a value of $|\mathbf{m}(O)|$.

\section{CONCLUSION}

Doping $La_2CuO_{4}$ with alkaline-earth, $Ae= Sr, Ba$, (and possibly co-doping with lanthanide $Ln = Nd, Eu$) generates holes in the $CuO_2$ planes of $La_{2-z-x}Ln_zAe_xCuO_{4}$ crystals. Pairs of the holes turn $O^{2-}$ ions to neutral oxygen atoms. A fraction of the atoms, $\tilde{O}$, of hole density $\check{p} \le  0.02$, is itinerant, skirmishing 3D-AFM, to suppression at $\check{p}$.
The remaining oxygen atoms, $O$, are stationary at anion lattice sites and form a superlattice that gives rise to both charge-order stripes and magnetization stripes with incommensurability $q_{c,m}(x) \propto \sqrt{x-\check{p}}$. The setoff value $\check{p}$ depends on the doping level and on temperature. Near $T = 0$ its value is $\check{p} = x_{N0} = 0.02$ for low doping, $x < 0.09$, but less, $\check{p} \approx 0.015$, in the doping range above.
With increasing temperature, however, the setoff value approaches the N\'{e}el concentration, $\check{p} \rightarrow x_N(T)$, at any doping level $x$. 
The antiparallel orientation of neighboring magnetic moments $\mathbf{m}(O)$ (with respect to the $O$ superlattice) yields a natural explanation for the coupling of $q_m(x) = \frac{1}{2} q_c(x)$, valid below the threshold of temperature dependence, $T < T'$.

Increasing density of the doped holes, hosted pair-wise in lattice-defect $O$ atoms of the $CuO_2$ planes, raises their chemical potential.
When doping exceeds a watershed value, $x > \hat{x}$, additional holes overflow to the $LaO$ layers where they also reside pair-wise in $O$ atoms. This leaves charge-order stripes of \emph{constant} $q_c$ in the $CuO_2$ planes.
The watershed concentration of the stripes' incommensurability, due to commencing hole population in the $LaO$ layers, may also cause a watershed division of the cooling curves of the intensity of X-rays, diffracted by stripes, when the crystals are cooled below the superconducting transition, $T < T_c$.

Doping $Ln_2CuO_{4}$ ($Ln$ = $Nd, Pr, Sm, La$) with $Ce$ generates electrons that occupy pairwise copper atoms in the $CuO_2$ planes of $Ln_{2-x}Ce_xCuO_{4}$ crystals, $Cu^{2+} + 2e^- \rightarrow Cu$. The large size of the $Cu$ atoms may cause the T $\rightarrow$ T' transition of `214' lattice structure at doping $x_6$. The magnetic moments, $\mathbf{m}(Cu) \; [\simeq \mathbf{m}(Cu^{2+})]$, align with the AFM of the host. This causes the high 3D-AFM stability, $x_{N0} \approx 0.13$, as well as the lack of magnetization stripes. The charge-order stripes have a square-root doping dependence, $q_c(x) \propto \sqrt{x}$, similar to the hole-doped `214' compounds.

Above the threshold temperature, $T > T'$, electron-hole pairs are thermally generated. They separate to reside pairwise at $Cu$ and $O$ atoms. The latter, adding to the $Ae$-generated holes, account for the increase of the incommensurability of charge order stripes with temperature. By aligning with the AFM of the host, the magnetic moments of thermally generated $Cu$ atoms counteract the magnetic moments from the $O$ atoms. In consequence, the locking of the incommensurability of charge order and magnetization stripes ceases, $q_m(x) \ne \frac{1}{2} q_c(x)$. The degree of antiparallel competition is determined by the relative magnitude of magnetic moments, $r = |\mathbf{m}(Cu)| \; \textbf{/} \;|\mathbf{m}(O)|$. If $|\mathbf{m}(O)| \approx \frac{1}{2} |\mathbf{m}(Cu^{2+})|$ can be found, then this would explain the observed \textit{decrease} of $2q_m(x,T > T')$ by about the same rate as the increase of $q_c(x,T > T')$.

\appendix

\section{DOPING DEPENDENCE OF CHARGE ORDER}

In the doping range from $x_6 = 2/6^2 = 0.056$ to the watershed value $\hat{x}$, the observed charge order incommensurability 
parallel to the planar axes,
\begin{equation} 
q^p_c(x)\Big|_{CuO_2} = \frac{\sqrt{2}}{2} \sqrt{x - \check{p}}\;,\;\;\;\;\;x_6 \le x \le \hat{x}  \;,
\end{equation}
can be derived with three assumptions. (i) A part $\check{p}$ of the doped holes in the $CuO_2$ plane is necessary to suppress 3D-AFM. (ii) The remaining part, $x - \check{p}$, forms a square superlattice parallel to the planar axes with (iii) \emph{two} holes per area $L_c(x)^2$,
\begin{equation} 
x - \check{p} = \frac{2}{L_c(x)^2}  \;.
\end{equation}
This gives the spacing of the superlattice as
\begin{equation} 
L_c(x) = \sqrt{\frac{2}{x - \check{p}}} \;.
\end{equation}
Charge-order incommensurability is defined as $q_c(x) \equiv 1/L_c(x)$. Inverting Eq. (A3) gives Eq. (A1).
An antiferromagnetic pattern, neighboring $L_c(x)^2$ areas harbor alternately oriented magnetic moments $\mathbf{m}$, so that the spacing between equally oriented moments is $2L_c(x)$.
This gives the incommensurability of magnetization stripes parallel to the planar axes as
\begin{equation} 
q^p_m(x)\Big|_{CuO_2} = \frac{\sqrt{2}}{4} \sqrt{x - \check{p}}\;,\;\;\;\;\;x_6 \le x \le \hat{x}  \;.
\end{equation}
At $Ae$ doping $x < x_6$, {\it diagonal} magnetization stripes are observed (but no charge order stripes). For the derivation of their incommensurability we partition the $CuO_2$ plane by diagonal squares of edge length $L'(x) = \sqrt{2} L_c(x)$. With this partitioning each $L'(x)^2$ square holds four holes,
\begin{equation} 
x - \check{x} = \frac{4}{L'(x)^2}  \;.
\end{equation}
Accounting for alternately oriented magnetic moments along the diagonal directions gives
\begin{equation} 
q^p_m(x)\Big|_{CuO_2} = \frac{1}{4} \sqrt{x - \check{p}}\;,\;\;\;\;\;x < x_6   \;.
\end{equation}

\section{RECALIBRATION OF A DOPING VALUE}

In the series of  stripe incommensurabilities of $La_{2-x}Sr_{x}CuO_{4}$, measured by Wen \textit{et al.},\cite{12} 
the value at $x=0.13$ nominal $Sr$-doping, $q_c=0.2281$, is considerably smaller than $q_c(x) = 0.235$, obtained from Eq. (2) for $x \ge 0.125$. Therefore a recalibration to an effective, stripe-generating concentration $x_{Sr}^{eff}$ is carried out with the following procedure: Data from  
the doping range where stripe incommensurability is constant, $q_c(x) = 0.235$, $x > 0.125$, are
used from the series under consideration\cite{12} and from a control group\cite{14} in order to obtain 
with Eq. (2) the corresponding offset value $\check{p}$. The $\check{p}$ value is then used to calculate with
Eq. (1) $q_c(x)$ for $x\le0.12$. The results are in close agreement with the experimental values (see Table III), which validates the procedure. The setoff $\check{p}=0.017$ and the experimental $q_c(0.13)=0.2281$ are then used to calculated with Eq. (1) an effective, stripe-generating $Sr$ doping, $x_{Sr}^{eff} = 0.121$.
\medskip 
\begin{table}[ht]
\begin{tabular}{|p{0.7cm}|p{1cm}|p{1.5cm}|p{1.7cm}|p{3.7cm}|p{4cm}|}
 \hline  \hline
Ref.&$T$ [K]  &  $\;\;\;x \;\;\;\;\;$ nominal &$q_c$ expt.  [r.l.u.] 
&$\; q_c \;$  calculated with Eq. (1) and $\check{p}= 0.013$   
&$\; q_c \;\;\;$  calculated with Eq. (1) and $\check{p} = 0.017$  
\\
 \hline  \hline
12&$23.3$  & $0.115$  & $0.226$ & $\leftarrow \;\;\;\leftarrow \;\;\;\leftarrow \;\;\;\leftarrow \;\;\;\leftarrow \;\;\;$ & $0.221$  almost agreement  \\
12&$28.4$  & 0.12  & 0.2264 & $\leftarrow \;\;\leftarrow \;\;\leftarrow \;\;\leftarrow \;\;\leftarrow \;\;\leftarrow \;\; $ & $\mathbf{0.227}$  close agreement \\
14&$40$  & 0.12  & 0.232 & $\mathbf{0.232}$ full agreement &   \\
\hline
12&$30.8$ & 0.13 & $\mathit{0.2281}\rightarrow$ & $\rightarrow$ used to recalibrate  &  with $\check{p} = 0.017$ \\

\hline
12&$37.5$  & 0.144  & 0.2321 &  & $\overline{0.23215}$ used in Eq. (2)\\
12&$35.5$  & 0.16  & 0.2322 &  & $\overline{0.23215}$ to obtain $\check{p} \; \uparrow$ \\
14&$40$  & 0.17  & 0.236 & $\overline{0.237}$ used in Eq. (2) &  \\
14&$40$  & 0.21  & 0.238 & $\overline{0.237}$ to obtain $\check{p} \; \uparrow$ & \\
 \hline   \hline
\end{tabular}
\caption{Incommensurability $q_c(x)$ of $La_{2-x}Sr_{x}CuO_{4}$ at temperature $T$, nominally doped with $Sr$ concentration $x$. From each research group the average experimental value in the $0.144 \le x \le 0.21$ doping range of constant $q_c$ is used (overlined in the bottom part of the table) to obtain offset values $\check{p}$ for calculating $q_c$ at $x \le 0.12$. It yields agreement (bold print) or almost agreement with experiment (top part of the table). 
With $\check{p}=0.017$ and $q_c=0.2281$ 
an effective, stripe-generating $Sr$ doping is calculated, $x_{Sr}^{eff} = 0.121$, instead of nominal $x=0.13$ (center line).}
\label{table:3}
\end{table}
\bigskip 

\centerline{ \textbf{ACKNOWLEDGMENT}}
\noindent I thank Jun-Sik Lee for providing $q_{cdw}$ data of $La_{2-x}Sr_xCuO_4$ and for helpful correspondence. Thanks also to Mark Dean for literature references.

\pagebreak


\begin{thebibliography}{68}
\bibitem{47} The data from Refs. 44 and 46 were used with a $Ba$ concentration re-calibrated according to Ref. 43.
\end{thebibliography}
\end{document}